\documentclass[aps,prl,twocolumn,showpacs,amssymb]{revtex4}

\usepackage[dvips]{graphicx}
\usepackage{dcolumn}
\usepackage{bm}
\def\be{\begin{equation}}
\def\en{\end{equation}}

\def\gdot{\dot{\gamma}}

\def\ls{\lesssim}
\newcommand{\bi}[1]{\mbox{\boldmath$#1$}}

\def\bea{\begin{eqnarray}}
\def\ena{\end{eqnarray}}

\begin{document}


\title{Plastic flow in polycrystal states  
in a binary mixture }


\author{Toshiyuki Hamanaka}
\affiliation{Department of Physics, Kyoto University, Kyoto 606-8502,
Japan}

\author{Akira Onuki}
\affiliation{Department of Physics, Kyoto University, Kyoto 606-8502,
Japan}


\date{\today}

\begin{abstract}
Using molecular dynamics simulation 
we examine dynamics in sheared polycrystal states 
in  a  binary mixture containing $10\%$ 
larger  particles in two dimensions.  
We   find 
large stress fluctuations 
arising from sliding motions 
of the particles  at the grain boundaries, which 
occur cooperatively  to release 
the elastic energy stored.
These dynamic processes are visualized 
with the aid of  a sixfold  angle  
$\alpha_j(t)$ representing the local crystal 
orientation and     a disorder variable    
$D_j(t)$ representing  a deviation 
from the hexagonal order for  particle $j$.  
\end{abstract}


\maketitle


Understanding 
the  deformation mechanisms  of 
polycrystals  under applied 
strain is of great scientific and technological 
importance in materials science \cite{Yip,Sch,Yama,Swy}. 
If the typical grain size exceeds 
a critical size $d_c\sim 10$nm, 
interplay of grain boundaries 
and  dislocation motions determines the mechanical properties. 
For very small grain sizes less  than $d_c$, 
some  simulations suggested that 
 plastic deformations 
are caused by sliding motions 
at the grain boundaries. 
There remain a number of puzzles 
in polycrystal rheology not 
yet simulated microscopically, 
such as the Portevin-Le Chatelier 
effect in dilute  alloys caused by intermittent yielding 
 \cite{Kubin}. 
On the other hand, in physics, dynamics  of sheared  glassy 
materials has been studied extensively 
 \cite{Liu,Onukibook}, 
including microscopic particle systems 
 \cite{yo,Miyazaki,Barrat}, 
granular materials \cite{Behringer,granular2}, and 
foams \cite{Okuzono}, while 
 not enough attention has  been paid to  
polycrystal rheology.

Recently,  using molecular dynamics simulation, 
we have examined the dynamics of 
polycrystal states realized on very small scales 
 in a model 
binary mixture \cite{HamaOnuki}.  
The parameters we have changed   are  the size ratio of 
the diameters of the two components $\sigma_2/\sigma_1$,  
the temperature $T$, and 
the composition $c$.  Polycrystal states  
  appear as intermediate states  
between crystal and glass,  
where  the grain boundary motions are severely slowed down 
with size dispersity.  
The particles in the grain boundary regions 
have a relatively high mobility 
leading to  
dynamic heterogeneity  
on long time scales. 
Thus investigating the jamming 
dynamics over wide ranges of 
$\sigma_2/\sigma_1$  and $c$ helps us  to 
 understand the glass dynamics as  highly frustrated limits. 
In this paper, we will  present 
  simulation results  
on polycrystal rheology, however,  at fixed 
$\sigma_2/\sigma_1$  and $c$.

Our two-dimensional (2D)  system is composed 
of the bulk region with volume $ V=L^2$ 
and the top and bottom  boundary  regions  with volume 
$0.1 V$, as can be seen in Fig. 1.  
Shear flow was realized by 
 the relative boundary motion of the boundaries, 
where the top and bottom velocities are 
 $\pm \gdot L/2$,  with $\gdot$ being an  applied 
shear rate.
In the bulk region, 
$0<x, y<L$, 
a mixture of large and small particles interact  
via a  truncated Lenard-Jones potential  
of the form,  $v_{\alpha\beta} (r)= 
4\epsilon [ ({\sigma_{\alpha\beta}}/{r})^{12}
-({\sigma_{\alpha\beta}}/{r})^{6}] -C_{\alpha\beta}  
$  ($\alpha, \beta=1,2$),  characterized by 
the energy  $\epsilon$   and 
the  soft-core diameter    $\sigma_{\alpha\beta}
=(\sigma_{\alpha}+\sigma_{\beta})/2$ with 
$\sigma_2/\sigma_1=1.4$. 
For  $r>r_{\rm cut}=3.2\sigma_1$, 
we set $v_{\alpha\beta} (r)=0$  
and the constant $C_{\alpha\beta}$  ensures  
the continuity of $v_{\alpha\beta}(r)$ at the cut-off 
$r=r_{\rm cut}$. 
In this paper, the particle numbers are  fixed  at  $ 
N_1=900$ and $N_2=100$ or $c=N_2/(N_1+N_2)=0.1$ in the bulk.  
The  volume $V$ is  
chosen  such that the volume fraction 
of the soft-core regions is  fixed at 
$1$ or at 
$\phi=  (N_1\sigma_1^2+N_2\sigma_2^2)/V=1$, so $L=33.1\sigma_1$. 
To each boundary ($-0.1L<y<0$ or $L<y<1.1L$),   
100 small particles with radius $\sigma_1$ 
are attached by the spring potential $10\epsilon |{\bi r}-{\bi R}_j|^2$. 
They also interact with the other particles 
in the boundary and bulk regions 
 with the common  Lenard-Jones potential. 
Before our  simulation, 
the attached positions  ${\bi R}_j$ ($j=1, \cdots,100$) in 
each boundary wall  were  
determined  in a liquid state realized  at $T=2\epsilon/k_B$ 
with the Lenard-Jones potential only.

We  integrated  the Newton   
equations using  the leapfrog algorithm   \cite{hansen}
under the periodic boundary 
condition  in the horizontal ($y$) direction, 
with the  mass ratio being  
 $m_{1}/m_{2}=(\sigma_{1}/\sigma_{2})^{2}$. 
The   time step 
of integration is  $0.002\tau$ with 
\be 
\tau=\sigma_{1}\sqrt{m_{1}/\epsilon}.
\en   
We will measure the time $t$ 
in units of $\tau$ and the shear rate $\gdot$ 
in units of $\tau^{-1}$. Without shear ($\gdot=0$), 
(i) we first equilibrated the system  in a liquid state at 
$T=2 \epsilon/k_B$  in a time interval of  $10^3$  
and then quenched it to the  final  temperature 
 $T=0.2 \epsilon/k_B$. 
(ii) After  a relaxation time of  
$5\times 10^3$, there was  no appreciable time evolution 
in various  quantities obtained as an average over the 
particles (see Fig.7 of Ref.\cite{HamaOnuki}). 
(iii) After these steps, we  applied a constant shear to the system.


\begin{figure}[b]
\includegraphics[scale=0.39]{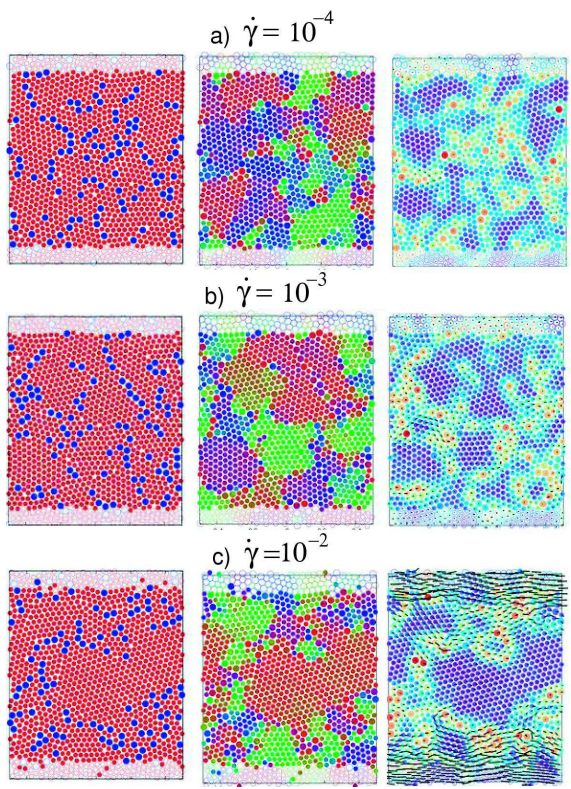}
\caption{(Color on line) 
Particle configuration (left), sixfold 
orientation order  (middle), and disorder 
variable (right) for three  sheared states, where  
$\dot{\gamma}= 10^{-4}$ (top), $10^{-3}$ (middle), 
and $10^{-2}$ (bottom),  in a binary mixture with 
$c=0.1$ and $\sigma_2/\sigma_1=1.4$. The snapshots for 
$\dot{\gamma}= 10^{-4}$ and  $10^{-3}$ are very similar, 
while for  $\gdot=10^{-2}$ 
the displacement vectors 
of the particles in a time interval of 10 
(right) show shear localization near the boundaries.
}
\end{figure}

Since a large fraction of 
the particles are enclosed by six particles 
in 2D dense particle systems, the  local crystalline order 
is  represented  by the  sixfold 
orientation \cite{NelsonTEXT}. 
We  define the orientation angle 
 $\alpha_j$ in the range $[0,\pi/3]$ 
 for each particle $j$ using the following complex number,  
\be 
\Psi_j=  \sum_{k\in {\rm bonded}}\exp[6i\theta_{jk}]=
|\Psi_j| e^{6i\alpha_{j}},  
\en   
where the summation is over  the particles  
''bonded''  to the particle $j$.  The 
two particles $j \in\alpha$ and 
$k \in\beta$ are bonded,  if  their  distance 
$ |{\bi r}_j-{\bi r}_k|$ is shorter than 
$R_{\alpha\beta}=1.25\sigma_{\alpha\beta}$ \cite{yo}. 
The upper cut-off $R_{\alpha\beta}$  is slightly 
longer than the first peak position of 
the pair-correlation function $g_{\alpha\beta}(r)$.  
The  $\theta_{jk}$ is the angle 
 of the relative vector 
${\bi r}_j-{\bi r}_k$ with respect to the $x$ axis. 
Next we construct  another  nonnegative-definite 
 variable representing 
the degree of disorder  for each particle $j$ by
 \cite{HamaOnuki}     
\be
D_{j} =    2\sum_{k\in {\rm bonded}}
 [1-\cos 6(\alpha_j-\alpha_k)]. 
\en 
For  a  perfect  crystal at low $T$ 
this quantity arises from  the  thermal vibrations  
 and is nearly zero, 
but  for particles  around defects it assumes large values 
in the range    $5-20$.

In Fig. 1, we show the particle configuration 
with the large particles in blue, the angles 
$\alpha_j$, and the disorder 
variable $D_j$  for   sheared states with   
(a) $\dot{\gamma}= 10^{-4}$,  (b)  $10^{-3}$, 
and (c) $10^{-2}$. See  Ref.\cite{HamaOnuki} 
for the color map of  $\alpha_j$ and 
Fig.3 below for that of $D_j$. 
The polycrystal grains and 
the grain boundaries are unambiguously  visualized 
using $\alpha_j$ and  $D_j$. 
By comparing the three kinds of snapshots for each 
state,  we notice that the large particles tend to 
form  the grain boundaries. 
This tendency can be seen even  without shear 
\cite{HamaOnuki}, 
but is  intensified under shear. 
Remarkably, the grain structures are 
insensitive   to  $\gdot$ 
for not very large shear ($\gdot \ls 10^{-3}$ here), 
where the effect of the boundary walls 
does not  extend into the bulk  
and the time-average of the horizontal  velocity is linear with the 
gradient $\gdot$.   To support this 
weakness of the boundary effect, 
almost the same grain structures were 
realized  under the Lee-Edwards 
boundary condition \cite{hansen} (not shown in this paper). 
However,  for very large shear $\gdot=10^{-2}$ in (c), 
the velocity gradient becomes localized 
near the boundaries  with larger  crystalline 
regions  in the middle being continuously rotated and 
deformed.

In Fig. 2,  the average shear stress 
$\sigma_{xy}(t)$ is displayed 
 in units of $\epsilon \sigma_1^{-2}$ as a function of  
the average strain $\gdot t$ after application 
of shear at $t=0$,   which is the 
sum of  the microscopic shear-stress contributions 
over all the  particles 
 in the bulk divided by $V$ \cite{Onukibook,hansen}.  
Each curve is a result of a single simulation run. 
It undergoes large temporal fluctuations, 
 arising from 
plastic deformations of the polycrystal structures. 
For not large shear,  it varies from a minimum 
about 0.4 to a maximum about 1.0  
on a   time scale  of order 
$0.1\gdot^{-1}$, and 
its time-average is insensitive to $\gdot$. 
The effective viscosity $\eta_{\rm eff}$ 
(time-average of $  
\sigma_{xy}/\gdot$)  thus behaves as $\gdot^{-1}$. 
At the largest shear $\gdot=10^{-2}$, 
the   time scale of the fluctuations 
is longer than $0.1\gdot^{-1}$ and 
 $\sigma_{xy}(t)$ is 
 larger than in the lower shear cases.

\begin{figure}[t]
\includegraphics[scale=0.44]{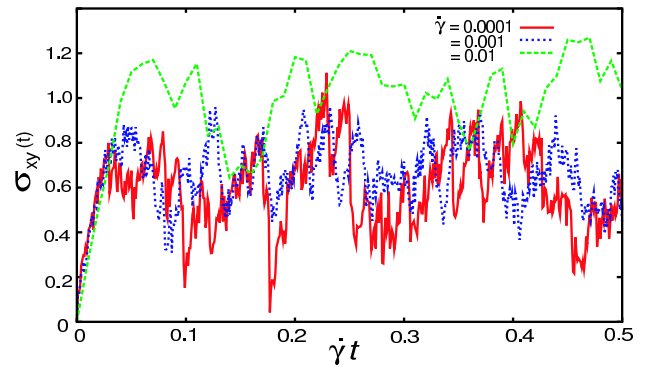}
\caption{(Color on line) Shear stress 
$\sigma_{xy}(t)$ vs strain $\gdot  t$ after application of shear 
 in units of $\epsilon \sigma_1^{-2}$ 
with $\gdot=10^{-4}$, $10^{-3}$, and $10^{-2}$, exhibiting 
large temporal fluctuations and 
strong  shear thinning.}
\end{figure}

\begin{figure}[ht]
\includegraphics[scale=0.4]{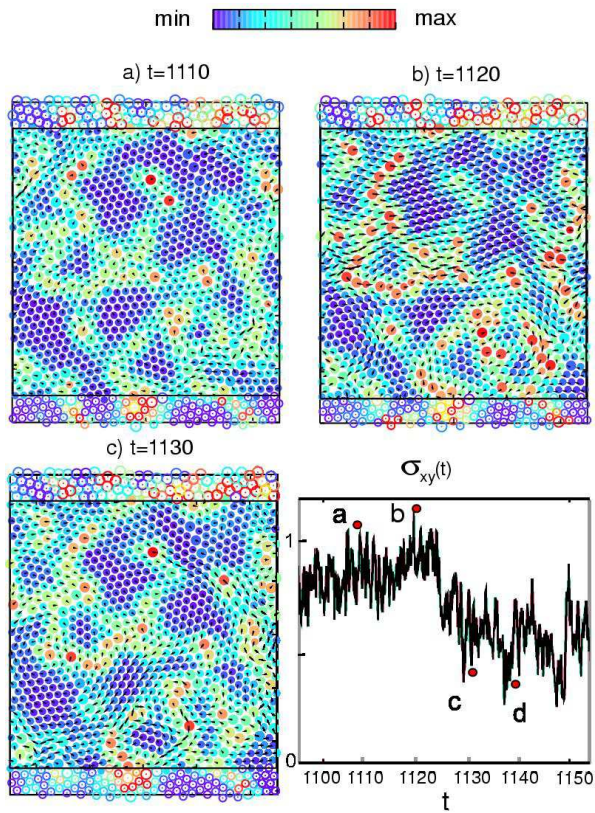}
\caption{(Color on line) 
Disorder variable $D_j$   
at (a) $t=1110$,  (b)  $1120$, and  (c) $1130$ 
at $\gdot=10^{-3}$. 
The arrows represent the 
 particle displacement $\Delta {\bi r}_j$ 
in the subsequent time interval of width  $10$, 
which are large in (b) for the particles in the grain boundary 
regions. The average stress $\sigma_{xy}(t)$ is also shown 
in this time region. 
}
\label{rot}
\end{figure}

In Fig. 3, 
we  display $D_j(t)$ at (a) $t=1110$, (b) $1120$, and 
(c) $1130$  at $\gdot=10^{-3}$,  superimposing  the 
displacement vectors $\Delta {\bi r}_j(t) 
= {\bi r}_j(t+\Delta t)-{\bi r}_j(t)$ 
with $\Delta t=10$.  The color 
is given  to each picture independently,  
according to its minimum 
and maximum of $D_j$.  
We also present  $\sigma_{xy}(t)$ 
in the corresponding time range. 
Between $t=1110$ and $1120$, the deformations are 
mostly "elastic" as in  (a) 
and  $\sigma_{xy}(t)$ gradually increases  
on the average. However, between $t=1120$ and $1130$, 
the picture  (b) indicates  significant   "sliding"
particle motions  in the grain boundary regions, 
which are of order $\sigma_1$ even 
for this small $\Delta t$. 
These sliding motions are triggered  collectively 
throughout the system (in our small system), 
leading to a catastrophic drop  of $\sigma_{xy}(t)$. 
The particles writtten in orange,  which are mostly larger ones, 
may be regarded to be in disordered configurations. 
Their number is of order  10 in (a) and (c), 
while it is about 50 in (b). 
In  (c)  noticeable particle displacements 
still continue.  Large scale collective motions  of 
the particles within  the grains 
 are  also conspicuous, which was already noticed in our 
previous simulation without shear \cite{HamaOnuki}.

In summary, in polycrystal states with very small 
grains, we have found intermittent yielding 
on a time scale of $0.1\gdot^{-1}$ for not very large 
$\gdot$.  It is  caused by 
cooperative  sliding  motions 
in the grain boundary regions   
in agreement with the  atomistic 
 simulations \cite{Sch,Yip,Yama,Swy}. 
In our small-scale simulation, however, 
we cannot determine 
the  spatial scale of the cooperative sliding extending over grains, 
which should be relevant in real systems.  
With increasing $c$, 
the typical sizes of the crystalline regions 
become smaller   and 
the stress fluctuations 
gradually   decrease. 
That is, weaker  disorder 
results in larger stress fluctuations in plastic flow. 
The proportionality 
of the structural relaxation  time to
$\gdot^{-1}$ and the shear thinning behavior 
$\eta_{\rm eff}\propto \gdot^{-1}$ 
still hold  for glasses,  
as observed experimentally \cite{Simmons} 
and   numerically  \cite{yo}. 
Similar plastic flow phenomena 
should be observable  in colloidal mixtures 
on expanded scales, as a  future experimental system. 
It is   of great  interest how our findings can be related 
to the other well-known  examples 
of enhanced stress fluctuations, 
where the effect of the size dispercity 
should be further examined   
\cite{Kubin,Behringer,granular2,Okuzono}.

\begin{acknowledgments}
This work was  supported by 
Grants in Aid for Scientific 
Research 
and for the 21st Century COE project 
(Center for Diversity and Universality in Physics)
 from the Ministry of Education, 
Culture, Sports, Science and Technology of Japan.
\end{acknowledgments}


\end{document}